\documentclass[runningheads]{llncs}

\usepackage[T1]{fontenc}
\usepackage{graphicx}
\usepackage{amssymb}
\usepackage{amsmath}
\usepackage{bbold}
\usepackage{color}
\usepackage{caption}
\usepackage{xfrac}
\usepackage{hyperref}
\hypersetup{
    breaklinks=true,
}
\usepackage[anythingbreaks]{breakurl}

\renewcommand{\paragraph}[1]{\textbf{#1}}
\usepackage{multicol, multirow}
\usepackage{booktabs}
\usepackage[margin=1in]{geometry}
\begin{document}

\title{Strategic bid response under automated market power mitigation in electricity markets}
\titlerunning{Strategic bid response under AMP}

\author{Chiara Fusar Bassini\inst{1,2,*} \and
Jacqueline Adelowo\inst{3,4} \and
Priya L.~Donti\inst{5} \and
Lynn H. Kaack\inst{1,2}}

%Chiara: \orcidID{0009-0004-2346-9132}
%Lynn: \orcidID{0000-0003-3630-3102}

\authorrunning{Fusar Bassini et al.}
\institute{Centre for Sustainability, Hertie School, Friedrichstraße 180, 10117 Berlin, Germany 
\and Data Science Lab, Hertie School, Friedrichstraße 180, 10117 Berlin, Germany
\and ifo Institut – Leibniz-Institut für Wirtschaftsforschung, Poschingerstraße 5, 81679 München, Germany
\and Ludwig-Maximilians-Universität München, Geschwister-Scholl-Platz 1, 80539 München, Germany
\and Massachusetts Institute of Technology, 77 Massachusetts Avenue, Cambridge, MA, USA
}

\maketitle              % % maximum 250 words
\abstract{In auction markets that are prone to market power abuse, preventive mitigation of bid prices can be applied through automated mitigation procedures (AMP). Despite the widespread application of AMP in US electricity markets, there exists scarce evidence on how firms strategically react to such price-cap-and-penalty regulation: when the price cap rarely leads to penalty mitigation, it is difficult to distinguish whether AMP are an effective deterrent or simply too lax. We investigate their impact on the bids of generation firms, using 2019 data from the New York and New England electricity markets (NYISO, ISO-NE). We employ a regression discontinuity design, which exploits the fact that the price cap with penalty is only activated when a structural index (e.g.,~congestion, pivotality) exceeds a certain cutoff. By estimating the Local Average Treatment Effect (LATE) of screening activation, we can causally identify successful deterrence of anti-competitive behavior. Around 30-40\% of the analyzed bidders per market exhibit a significant strategic response -- corresponding to a decrease in maximum bid prices of 4-10 \$/MWh to avoid the penalty. However, there is significant heterogeneity between firms, and the regulatory impact on the overall market is not statistically detectable, suggesting lax mitigation thresholds. Using a merit-order simulation, we estimate the welfare impact of more stringent thresholds to lie between 350 and 980 thousand dollars of increased buyer surplus per mitigated hour, with the associated number of mitigated hours being below 33 hours/year. Our results motivate the empirical calibration of mitigation thresholds to improve the efficiency of AMP regulation.}
\newline\newline
\noindent $^*$ Corresponding author: \texttt{c.fusarbassini@hertie-school.org}
\newpage
\section{Introduction}
\begin{figure}[htpb]
    \centering
\includegraphics[width=0.85\linewidth]{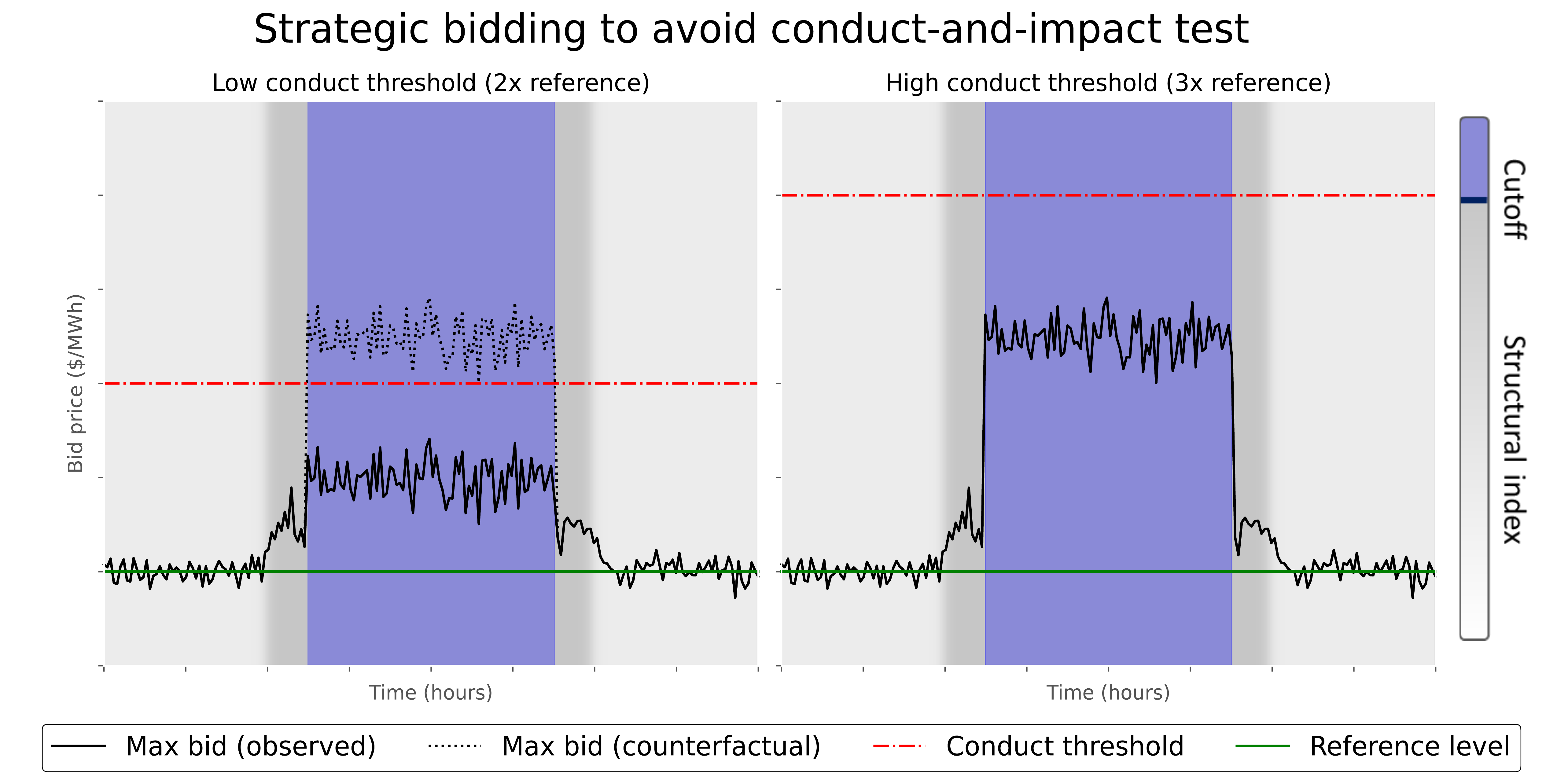}
\caption{Strategic bid price adjustment under AMP. If the structural index exceeds a cutoff, screening activates. With a tight conduct threshold (left), the unit submits lower prices to avoid mitigation; a loose one (right) has no effect on bids. Both cases result in no mitigation: the left one, because AMP is effective; the right one, because it is not.}
    \label{fig:example}

\end{figure}

In electricity markets, the physical constraints dictated by the grid structure, the limited short-term flexibility of demand, and the volatility of real-time prices often create time windows where there is both the incentive and the opportunity to exercise market power~\cite{xu}. In fact, the electricity sector has been characterized by several anti-competitive cases, spanning across countries and time periods; the most notorious example is probably the 2000 California electricity crisis, where manipulations led to price increases of as much as 500\%~\cite{joskow2002quantitative,borenstein2002measuring} and which triggered substantial effort to regulate electricity markets. Market power abuse is a market failure: non-competitive electricity prices affect virtually all consumers, leading to welfare losses, wealth transfers from buyers to sellers, and distorted investment signals. Regulatory bodies usually tackle market power abuse \textit{after} the incidents have occurred, by conducting ex-post mitigation of anti-competitive behavior, and collecting fines or damage payments. However, ex-post mitigation of anti-competitive bidding is often decided on a case-by-case basis and involves lengthy investigations, which make it inefficient and unpredictable~\cite{brattle}.
\newline \newline
To facilitate prompt and efficient mitigation of market power abuse, many electricity markets in the United States (US) employ ex-ante mitigation procedures, designed to increase transparency and streamline regulatory intervention. These automated mitigation procedures (AMP) involve real-time screening of wholesale auction markets to identify bid prices that suggest market power abuse~\footnote{These bids can be market-distortive in two ways: through a direct and excessive increase of clearing prices or through an indirect and artificial reduction in supply (economic withholding).} and automatic mitigation, i.e.,~overwriting of bid prices with (lower) competitive benchmarks. When the market is particularly sensitive to market power abuse, e.g.~due to pivotality or congestion, AMP enforce a temporary, unit-specific, dynamic price cap. If the bid prices of a generation unit exceed this conduct threshold in a market-distorting way, its bids are penalized down to a unit-specific competitive benchmark, its so-called reference level. Hence, AMP are a price-cap-and-penalty regulation. Both the specific conduct threshold and the reference level of a unit are computed on the basis of its generation costs and its historical bid prices. AMP thus make market power mitigation less arbitrary, as generation firms know that they will be mitigated during critical hours and are informed about their reference levels -- although they can still contest mitigation, e.g.,~if they can prove that their price caps are inaccurate due to higher procurement costs~\cite{nyiso_manual}. 
\newline \newline
AMP are rule-based, multi-step procedures and, as such, they heavily rely on the selected mitigation thresholds. Regulators monitor the competitiveness of the market using structural indices for pivotality or congestion. When the market is tight, i.e.,~the structural index surpasses a cutoff value, conduct thresholds are activated. These thresholds allow firms to exceed their reference levels (green line in Figure~\ref{fig:example}) up to a certain tolerated conduct threshold (red line) before enforcing the regulation and mitigating the bid down to its reference (again, green line). AMP are based on the notion of \textit{workable} competition, which recognizes that moderate markups on perfectly competitive prices are inevitable~\cite{kahn1988economics} and mitigation is applied based only on the suspected (but difficult to prove) abuse of market power. This creates a difficult task for regulators: choosing an adequate price cap, which tolerates moderate markups and prevents overmitigation to negative producer surplus, but is tight enough to deter market power abuse. Since their introduction in the early 2000s, AMP thresholds have never been changed, nor were they ever empirically validated~\cite{isone2022}. This raises the question whether they are set in an appropriate and efficient way. Because AMP are preventive measures against the exertion of market power, it proves difficult to assess their actual impact on the market, as their impact is reflected in how rarely bids need to be mitigated~\cite{goldman2004review}. In other words, because AMP are intended to deter generators from exercising market power, a lack of mitigated hours may actually signal effectiveness, insofar as firms refrain from raising prices in the first place.
 \newline \newline
 In this paper, we investigate the deterrent effect of the dynamic and temporary conduct thresholds of AMP and analyze whether they significantly impact the bidding behavior of generation firms. Our study addresses two main questions:
\begin{enumerate}

    \item Do firms strategically adjust their bids to avoid triggering penalty mitigation?\label{q1}
    \item How can the design of AMP regulation thresholds be improved from a welfare transfer perspective? \label{q2}
    
\end{enumerate}
To address the first research question, we leverage a regression discontinuity design (RDD), where we exploit the fact that AMP are activated only if structural indices for pivotality or congestion exceed a certain cutoff. In other words, the unit-specific, dynamic price caps are activated only during hours where the structural index is beyond a cutoff, creating a discontinuous treatment. Screening for adverse conduct becomes random at the margin around this cutoff of the structural index, %and screening for adverse behavior becomes random at the margin around this cutoff. 
allowing us to causally identify the Local Average Treatment Effect (LATE) of AMP activation on observed bid prices. Our RDD can distinguish between two possible bidding strategies, conceptually illustrated in Figure~\ref{fig:example}. The left panel displays a strategic adjustment of maximum bids in response to AMP screening: the maximum bid is strictly below the conduct threshold when AMP are activated, compared to the counterfactual without regulation. This would be consistent with a successful deterrent effect of the regulation. Instead, the illustration on the right with a looser threshold is an example of regulatory laxity and shows how bids could also \textit{not} change significantly during screening compared to the counterfactual. To address the second research question, we undertake a simulation-based selection of tighter conduct thresholds. Using a merit-order model, we simulate real-time prices under different variations of AMP, with more or less stringent mitigation thresholds.
\newline \newline
This paper provides novel guidance to evaluate and improve AMP design. Our contribution is two-fold: (1) To the best of our knowledge, we are the first to provide an empirical strategy to evaluate the effectiveness of ex-ante regulation through AMP, distinguishing between successful deterrence on the one hand, and lax, non-binding price caps on the other. Our causal methodology allows us to differentiate between two cases: AMP is not triggered because the procedure works as intended and firms strategically adjust their bidding behavior to avoid the penalty, or because the conduct thresholds are not binding. (2) We quantitatively estimate the welfare benefits of calibrating AMP thresholds, strengthening their efficiency while limiting market intervention. Regulators can use the simulation to set AMP thresholds designed either to achieve a welfare goal or to control how many hours are mitigated. We use one year of bid data (2019) from two different markets to demonstrate the empirical validity of our methodology: ISO New England (ISO-NE), the system operator for Connecticut, Maine, Massachusetts, New Hampshire, Rhode Island and Vermont, and NYISO, for the State of New York.
\newline \newline
Our empirical results shed light on the limited strategic reactions of firms in markets with AMP. We find that at the market-level, the point estimates of the bid price adjustment in response to AMP range at negative 1-2~\$MWh; however they are not statistically significant. Taking into account heterogeneous firm responses, we find that at the firm level, 30-40~\% of analyzed bidders do show a statistically significant response to AMP, corresponding to a median decrease of 4-5~\$/MWh for ISO-NE and 9-10~\$/MWh for NYISO. As we do not find a statistically significant market-wide response, we point out a major potential shortcoming of current AMP: lax conduct thresholds. ISO-NE itself reports that \emph{the current thresholds allow for considerable latitude in supply offers levels over competitive benchmarks [...] and have been in place for many years with little empirical support}~\cite{isone2022}. In absence of successful market-wide deterrence through AMP, we simulate how more stringent procedures can be implemented without excessively increasing the number of mitigated hours (33 hours per year, or less), thereby limiting market interference.
\newline \newline
It is equally important to note what our study will \emph{not} address. Because we only assess whether companies respond to the expectation of being mitigated, with our approach we are neither able to confirm nor to refute the existence of market power abuse in the analyzed markets. We do not address mitigation of other forms of market abuse beyond excessive bidding, focusing solely on two specific types of AMP which mitigate bids for incremental energy (i.e.,~\$ per MWh of additional generation). The mitigation of other manipulation strategies, such as the physical withholding of assets or tacit collusion between generation firms, is not within the scope of this paper~\footnote{For an overview of screening methods for collusion, see e.g.~\cite{Brown2023}.}.
\newline \newline
The remainder of the paper is organized as follows. Section~\ref{sec:literature} summarizes the literature and the existing AMP implementations. Section~\ref{sec:methodology} introduces the bid data and the empirical methodology. Sections~\ref{sec:results} and~\ref{sec:discussion} respectively present the results and discuss their implications and limitations, and Section~\ref{sec:conclusion} concludes.

\section{Background}\label{sec:literature}
\subsection{Previous work}
To quantify AMP the ability to abuse market power, most AMP use structural indices of market concentration. Due to idiosyncrasies of the electricity market, static structural indices such as the Herfindahl-Hirschman Index (HHI), which are common in other markets and measure the market power of a company based on its relative size (i.e.,~installed capacity) can over- or underestimate the market power of generation companies~\cite{newbery2009predicting}. More suitable, and therefore often used for AMP, is the Residual Supplier Index (RSI~\cite{sheffrin2002predicting}), a structural index measuring the pivotality of a supplier. The RSI is a continuous and dynamic index which quantifies the extent to which the market depends on a specific supplier to satisfy the current system demand: with an RSI lower than 1, demand cannot be covered without its supply. During hours of tight market supply, AMP screen the conduct of the generation company, i.e.,~whether its prices are considerably higher than a benchmark. Although behavioral indices, such as the Lerner Index, can offer a simple measure of price-cost margins, their application is limited by the difficulty of accurately determining production costs~\cite{twomey2005review}. Reference levels, which are a unit-specific proxy for competitive prices, are used instead.
\newline \newline
Previous research on preventive mitigation in electricity markets focused on improving the design of AMP procedures. For instance, Graf et al.~\cite{Graf2021} give an overview of the challenges related to existing AMP; Adelowo and Bohland~\cite{adelowo2024} propose and test strategies to reduce their manipulability and increase their precision and their welfare gains. Simoglou et al.~\cite{Simoglou2025} provide an extensive simulation of the European electricity system, where electricity prices in several EU markets are found to decrease if moderately tight AMP were to be applied. However, green-field, simulation-based studies on the mitigating impact of AMP on wholesale prices allow only limited insight into how AMP are received by real-world market participants, because they omit adjustment strategies of firms. Unfortunately, little empirical evidence has been produced on how implementing AMP affects the dynamics of electricity markets. Entriken and Wan~\cite{entriken2005agent} use an agent-based simulation that shows that AMP can reduce market clearing prices when bids would otherwise reach price caps. Two experimental studies with human participants find that generation firms can successfully manipulate AMP, for example by inflating reference levels~\cite{shawhan2011experimental,kiesling2007experimental}. However, the knowledge gained through these experiments was never tested on real-world data. The present study addresses this gap, using one year of unit-level bidding data from two US markets with long-established AMP in place.

\subsection{Regulatory context}
\begin{figure}[htbp]
    \centering
\includegraphics[width=0.85\linewidth]{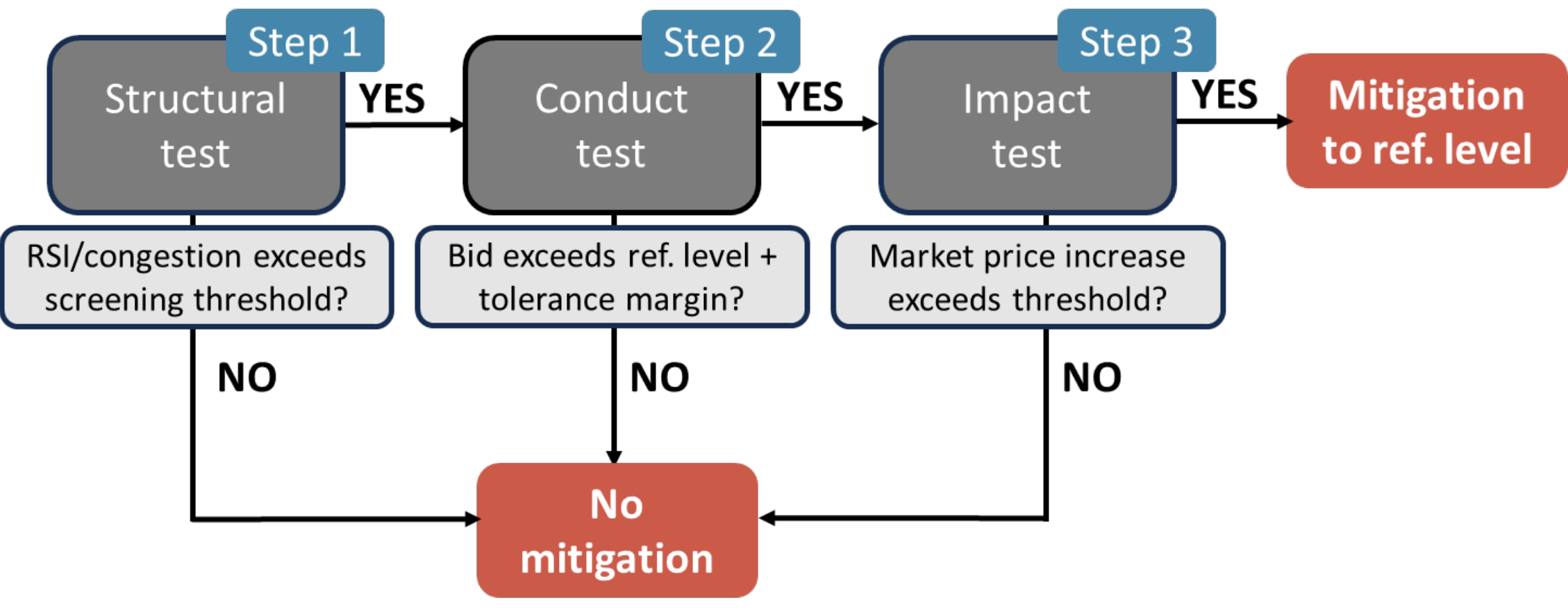}
    \caption{Structure of automated mitigation procedures.}
    \label{fig:amp_flowchart}
\end{figure}

In the late 1990s, the electricity sector in the US and Europe underwent extensive restructuring: in an effort to lower structural market power, vertically integrated monopolies were dismantled and competition between generation firms was enhanced. Liberalization reforms were driven by the expectation that competition would directly yield more efficiency and, ultimately, lower prices for consumers. However, the efficiency increase in electricity generation did not always translate into lower consumer prices, but instead into higher profit margins for producers~\cite{newbery1997privatisation}. In fact, without adequate regulatory oversight and appropriate market design, liberalization can fail to prevent the exercise of market power~\cite{joskow2008lessons}. Against this background, in 1999, the New York Independent System Operator (NYISO) introduced automated mitigation procedures to correct excessively high energy bids~\cite{peterson2001best}. In the aftermath of the California electricity crisis, other markets followed its example; as of today, six US system operators (ISO-NE, CAISO, MISO, NYISO, PJM and ERCOT) utilize AMP.
\newline \newline
The two main antitrust agencies in the US -- the Federal Trade Commission and the Federal Energy Regulatory Commission -- define market power as \textit{the ability to profitably maintain prices above competitive levels for a significant period of time}~\cite{brattle}. Among their proposed solutions to limit the abuse of market power, AMP stand out as the only ex-ante procedures currently implemented in wholesale electricity markets. AMP are integrated in the market-clearing software and can overwrite the offers of generation companies which are found to excessively influence the market outcome. Structural procedures apply mitigation when market conditions (pivotality, congestion, or reliability requirements) may grant a supplier with market power, while conduct-and-impact procedures first apply two sequential tests: a conduct test that compares the bids of a generation unit against its reference level (augmented by a conduct tolerance threshold), and an impact test that assesses the influence of these bids on the final market price~\footnote{We refer to~\cite{twomey2005review,Graf2021,adelowo2024} for detailed descriptions of AMP.}. A combination of the two approaches is possible: when structural conditions are met, a conduct-and-impact screening is applied, as shown in Figure~\ref{fig:amp_flowchart}. 
\newline \newline 
In the cases under analysis, a structural test (Step 1) tests whether the RSI of the bidding firm (ISO-NE) or congestion (NYISO) are beyond a certain cutoff. If so, conduct screening is activated (Step 2). Firms can now only bid up to their conduct threshold, which is equal to the reference level plus a certain tolerance; otherwise, they risk mitigation down to their reference level. Hence, the conduct threshold constitutes a temporary price cap. The reference level in this context represents a proxy for the competitive bid price of a unit and is typically derived from its historical accepted offers (offer-based), from its marginal cost estimates (cost-based), or from historical market prices (LMP-based). If the conduct test fails because a bid price exceeds the tolerance threshold, a subsequent impact test is performed to attest its impact on the market price (Step 3). If the impact is significant, bids are mitigated, i.e.,~substituted by their reference level. Note that the reference level to which bids are mitigated is always lower than the conduct tolerance threshold up to which the unit could have bid without triggering mitigation. Hence, mitigation includes a price penalty of the size of the tolerance margin; we therefore dub this type of policy a price-cap-and-penalty regulation. 
\section{Methodology} \label{sec:methodology}
\subsection{Data}
Our analysis centers around the electricity markets of New England (ISO-NE) and New York (NYISO) because the two markets share similar structures and implement comparable AMP regulation, as illustrated in Table~\ref{tab:market_characteristics}. Both employ conduct-and-impact mitigation, with more stringent mitigation thresholds in NYISO due to the higher market congestion. Both systems are similar in size and, as of 2019, 50-60\% of their installed capacity was gas-fired generation, making prices particularly sensitive to fluctuations in natural gas prices. Clearing prices in that period were also closely aligned at around 30 \$/MWh. The 2018–2019 period thus provides a suitable study window, characterized by market stability and the absence of significant structural shocks, hence allowing a clean analysis of the baseline response of firms. %We also report the number of hours during which the market experienced congestion or pivotality of at least one supplier, as these conditions favor market power abuse. Notably, NYISO exhibited double the congestion levels of ISO-NE; however, for NYISO, we could not compute firm-level RSI due to the incomplete mapping between bidders and firms. 
\newline \newline
ISO-NE and NYISO use nodal pricing, setting a different price of electricity at each point on the grid to reflect transmission constraints. Load zones, i.e.,~geographically defined aggregation of multiple transmission nodes, are used for pricing and settlement purposes. As usual for wholesale electricity markets, both ISOs have adopted a system of sequential auctions with uniform pricing. The day-ahead market clears electricity demand and supply one day in advance for the following day; here, bidders can submit hourly bids (price-quantity pairs) for each unit. After the closure of the day-ahead market (1:30 PM), the real-time market opens: demand forecasts are updated, and bidders can revise bids or add bids for units that were not scheduled in the day-ahead market. This real-time auction is the focus of our study, because it is the market where the analyzed AMP are applied. We obtain market data from the data platforms of the two ISOs; a full list of data sources is provided in Table~\ref{tab:sources}.
\newline \newline
We analyze hourly unit-level bid data from March $1^{\text{st}}$, 2018 to December $31^{\text{st}}$, 2019. Every real-time bid corresponds to a physical generator, but because the generation units are anonymized, their fuel type and location cannot be identified. Each bid includes a maximum quantity of supplied energy, start-up, incremental and operating costs for a single generation unit, as well as information on its current status, i.e.,~whether it is unavailable for generation, scheduled for must-run, or scheduled for regular economic dispatch. To allow for flexible bidding, each unit can submit an increasing step-function of price-quantity bids for incremental generation (up to 10 steps in New England, or up to 12 in New York). For our analysis, we focus on the maximum bid price, because if one of the incremental bids of a unit fails conduct-and-impact screening, \emph{all} its bids are mitigated, not only those above the conduct threshold. 
\subsection{Regression discontinuity design}\label{subsec:RD}
Table~\ref{tab:amp} summarizes the two AMP under consideration. Their multi-step nature provides a quasi-experimental setting around the structural cutoff (Fig.~\ref{fig:amp_flowchart}, Step 1) used to activate the conduct screening (Fig.~\ref{fig:amp_flowchart}, Step 2). We exploit a regression discontinuity design (RDD) to assess the impact of AMP activation on the maximum bid prices submitted in the real-time market. We can thus causally determine whether bidders alter their real-time behavior based on their expectation of being screened for mitigation. This hypothesis is supported by empirical evidence showing that firms are usually able to form ex-ante beliefs about the supply curves of their competitors, which allows them to bid strategically and in a profit-maximizing manner~\cite{Klemperer.1989,Green.1992}. In the New England market, the fact that this procedure is applied only in the real-time market, and not in the day-ahead, reinforces strategic behavior. It allows bidders in the market to use the clearing of the day-ahead to update their beliefs on whether they will be pivotal in the real-time market; and therefore screened for mitigation.
\newline \newline
The maximum bid price submitted by a bidder for a given hour and generation unit reflects whether the unit risks being mitigated. We can expect the maximum bid price to be endogenous to the current degree of market power of a bidder, as lower RSI and increased congestion (the running or score variable of the RDD) are associated with a firm's increasing ability to raise prices. Therefore, the maximum bid (outcome variable) is not exogenous to the screening status -- a binary treatment that denotes whether the score variable exceeds the cutoff value (Step 1), activating the conduct threshold (Step 2). However, our RDD exploits the fact that a decreasing RSI or increasing congestion indicate a \textit{continuous} increase in market power, while the screening is activated in a \textit{discontinuous} manner at the cutoff. In the immediate proximity of the structural cutoff, the marginal change in market power goes to zero, so that all differences in bidding behavior immediately left and right of the cutoff can be causally traced back to the activation of conduct screening. In other words, in the immediate vicinity of the structural cutoff, the conduct screening becomes random, and thus we can identify the local average treatment effect of AMP activation.
\newline\newline 
Depending on the side of the cutoff where the treatment is applied (to the left for pivotality, to the right for congestion), we define a shifted score variable $\tilde{S}$ centered at zero:
\begin{eqnarray}
\tilde{S}_{t,g} & = &
\begin{cases}
         c - S_{t,g} &  \text{if } T_{t,g} = \mathbb{1}\{S_{t,g} \leq c\}\\
         S_{t,g} - c & \text{if } T_{t,g} = \mathbb{1}\{S_{t,g} \geq c\},
    \end{cases} \label{eq:scorevariable} 
\end{eqnarray}
where indices $t$ and $g$ represent the hour and generation unit, $T$ the binary treatment status, $S$ the original score variable and $c$ its cutoff value. The baseline regression equation used to estimate the LATE is:
\begin{eqnarray} 
    p^{\text{max}}_{t,g} &=& \beta_0 + \beta_1 T_{t,g} +  \beta_2 \tilde{S}_{t,g} + \beta_3 \tilde{S}_{t,g}  \times T_{t,g} +\beta_4 \text{ref}_{t,g} + \beta_5 \text{gas}_t +  \epsilon_{t,g}, \label{eq:RDD_baseline}
\end{eqnarray}
where $p^{\text{max}}$ is the highest submitted bid price in \$/MWh and $\text{ref}_{t,g}$, $\text{gas}_{t}$ are the control variables (respectively, reference level and gas price).
%\subsubsection{AMP structure}\label{subsubsec:structural}
\newline \newline
We use the structural indices employed by the two AMP as the score variable $S$ and its cutoff value $c$ in the RDD. For ISO-NE, we use the Residual Supply Index (RSI) as the score variable with a cutoff of 1, where the system-wide RSI is measured as:
$$\text{RSI}_{t,f} = \frac{\text{Market Supply}_{t} - \text{Firm Supply}_{t,f}}{\text{Total Load}_{t} \;+ \;\text{Reserves}_{t}},$$
and a firm $f$ is considered pivotal for hour $t$ with an RSI lower than 1. In fact, if the RSI is less than one, the supplier firm is essential to meet system demand. Our calculation of this structural index follows the ISO-NE market manual~\cite{isone2019}, but excludes unavailable units and must-take energy from must-run units. Moreover, since firms do not observe realized demand when submitting bids, we use load forecasts as the denominator. Since our estimated share of hours with at least one pivotal supplier (around 12\%) is close to the estimate from ISO-NE internal market report (17\%, see Table~\ref{tab:market_characteristics}), we assume our calculation to be a good proxy of the RSI applied in practice. Bids submitted during pivotal hours are flagged and evaluated against a unit-specific conduct threshold and undergo an impact test, which evaluates whether such bids significantly impact the market. 
\newline \newline
In the case of NYISO, the structural test measures congestion. This test assesses whether the real-time shadow price at the unit node, which indicates the marginal cost of congestion, is greater than 0.04 \$/MWh. The New York City area is excluded from this calculation, as it is always considered congested. Within congested areas, the conduct threshold is computed based on the 12-month nodal price average and the frequency of congestion. For example, in the Capital area, which was congested for 187 hours in 2018 and had an average real-time locational marginal price of 37.98 \$/MWh, the threshold would be 35.58 \$/MWh. All units within the same area are subject to this cutoff, or their own unit-specific conduct threshold, whichever is lower. Since we lack detailed location data for generation units, we approximate market-level congestion using the market-level average shadow prices, weighted by the load of each load area $a$, with a lag of one hour: 
$$\text{Congestion}_{t} = \sum_{a}\frac{\text{Area Load}_{t-1,a}}{\text{Total Load}_{t-1}} \cdot \text{Shadow Price}_{t-1,a}.$$
The congestion indicator is lagged and load-averaged to address endogeneity of bids and prices and heterogeneity in the dimension of the load areas.  
\newline\newline
Crucially, we are not interested in the bid price itself, but in whether it falls above or below the conduct threshold of a unit. Since its conduct threshold linearly depends on the unit's reference level, we linearly control for it in the RDD. The reference level of a unit approximates its current competitive price: it is dynamic, because it changes with its bidding history and with market prices, and it is always known to the bidder, who can use it to calculate its conduct threshold. Due to data confidentiality, reference levels are not publicly available; following the offer-based method, we therefore calculated them using the weighted average of economic bids between 0 \$/MWh and 800 \$/MWh submitted by the asset in the last 90 days~\cite{adelowo2024}. Figure~\ref{fig:ref_level} shows the reference level for a sample unit, alongside its maximum bid. Finally, given that fuel price adjustments can affect reference levels and both ISOs heavily rely on gas generation, we also control for natural gas prices at the Henry Hub, the main price benchmark for the US market~\cite{gas}. 
\begin{table}[htbp]
    \centering
    \begin{tabular}{p{1.8cm}p{4.5cm}p{4.5cm}} \toprule
    & \textbf{ISO-NE} & \textbf{NYISO} \\\midrule
    \textbf{Structural index}& Market-level pivotality & Congestion  \\
    \textbf{Structural cutoff} & RSI $\leq 1$ & Shadow price \newline $\geq 0.04 \,\$/\text{MWh}$ \\
    \textbf{Conduct threshold}& Min of 100 \$/MWh and 300\% increase in ref. level& \underline{$2\% \times \text{avg.~price} \times 8760$}   \,$\text{ constrained hours}$ \\
    \textbf{Impact threshold} & Min of 100 \$/MWh and 200\% increase in LMP &  - \\
    \textbf{Market} & Real-time & Day-ahead, real-time\\\bottomrule\\
    \end{tabular}
    \caption{Analyzed AMP. The structural index serves as the score variable of the RDD; the treatment (AMP screening) is activated at the structural cutoff.}
    \label{tab:amp}
\end{table}
\subsection{Model specifications}\label{subsec:specifications}
To test the robustness of our model, we propose different specifications. We first estimate a pooled market-level regression for each of ISO-NE and NYISO. The pooled model is a linear regression that considers units from all bidders in the market. To account for bidder-specific strategies and related heteroskedasticity in the residuals, we extend Equation (\ref{eq:RDD_baseline}) to include bidder fixed-effects and cluster standard errors accordingly. To gain more insights into the heterogeneity of firm responses, we also estimate the RDD separately for each bidder, as their units are likely to follow similar strategies. Although the structural equation remains the same as in the market-level specification, we now account in a more sophisticated way for the differences in bidding strategies between firms by fitting one regression per bidder. Note that, strictly speaking, the term ``bidder'' refers in our case to an anonymized supplier bidding in the real-time market. For ISO-NE, this is directly equivalent to a generation firm, while in the case of NYISO multiple bidders can belong to the same generation firm, and vice versa~\cite{nyisobiddata}. 
\newline \newline
Empirically, RDDs are often limited by the compliance rate of a treatment, which reflects how closely the actual treatment assignment matches the rule defined by the cutoff in the score variable. In our case, bidders cannot ``opt out'' of AMP -- meaning that the treatment is directly imposed without an explicit compliance decision -- however, they are not informed by the ISO of screening activation. The degree to which bidders anticipate the treatment can therefore be regarded as an implicit measure of compliance: a bidder that can precisely predict when it is pivotal or when congestion occurs has the ability to perfectly respond to screening activation. If we assume perfect compliance, we obtain a sharp RDD. However, anecdotal evidence, including conversations with ISOs and market experts, suggests that even large, experienced firms cannot perfectly determine their pivotal status or the current congestion level. To account for this inherent treatment uncertainty, as well as possible measurement uncertainty in the structural index caused by noise in the data, we compare the results of the baseline sharp specification with a fuzzy specification. Full details on the fuzzy RDD implementation are provided in Appendix~\ref{sec:fuzzy}. 
\newline \newline
Finally, AMP activation is relatively rare, as most hours do not exhibit pivotal suppliers or strong market-level congestion. In both cases, the score variables are thus unevenly distributed left and right of the cutoff. In ISO-NE, the market operates mostly with sufficient excess capacity, so that there are fewer observations on the right-hand side of the cutoff. In NYISO, extremely (un-)congested events skew the distribution of the load-averaged shadow prices. These data points are far from the cutoff value and less informative about the discontinuity at the cutoff, which represents our treatment effect. Theory suggests that, when the parametric relationship between score and outcome variables is not known, constructing bandwidths for local estimation around the treatment cutoff is particularly robust~\cite{gelman2019high,huntington2021effect}. We therefore adopt a locally linear parametrization, and describe the procedure to select bandwidths in Appendix~\ref{sec:bandwidth}. In addition to the baseline sharp specification with a narrow bandwidth, we tested a wider specification to explore its impact on the estimated coefficients. The results of further robustness tests with different bandwidths and higher-order interactions are presented in Table~\ref{tab:sensitivity}.
\subsection{Merit-order simulation}\label{subsec:moc}
The AMP applied by ISO-NE rely on several thresholds that are historically motivated and have limited empirical justification~\cite{isone2022}. To investigate how the effectiveness of the regulation could be improved, we construct a simulation of ISO-NE’s real-time market clearing using actual market data. Each scenario simulates market-level real-time prices with hourly granularity, based on 2019 market-level demand and incremental bids of available generation units. To evaluate the impact of altering mitigation thresholds, the simulated prices are compared to a simulated baseline scenario, where the current AMP thresholds are used. 
\newline \newline
For a given time hour $t$, a generation unit $g$ in the set of currently available units $G_t$ can bid up to 10 incremental bids; each bid $b$ is composed of price $p_{g,b,t}$ and quantity $q_{g,b,t}$. The model selects the set of incremental bids which minimize the total cost of incremental energy and satisfy demand by solving the integer linear problem: 
\begin{align*}
  &\min_t \;f(t):= \sum_{g\in \mathcal{G}_t} \sum_{b=1}^{10} x_{g,b} \cdot p_{t,g,b} \cdot q_{t,g,b} \\
 & \text{subject to:}\\
 &  \sum_{g\in \mathcal{G}_t} \sum_{b=1}^{10} x_{g,b} \cdot p_{t,g,b} \geq L_t\\
 &x_{g,b} \in \{0,1\} \hspace{3cm} \forall g \in \mathcal{G}_t, b \in \{0, ...,10\},
\end{align*}
where $x_{g,b}$ is a binary variable that represents whether a given bid is selected and $L_t$ is the total load forecast. The resulting $P^*_t := \arg \min_t f(t)$ is the market-level clearing price. In a second step, the mitigation procedure presented in Table~\ref{tab:amp} is applied to bid prices $p_{t,g,b}$: here, each of the simulated scenarios applies different AMP thresholds. If, for a given hour, AMP fails, we compute a new, mitigated clearing price for the market, $P^m_t$. For each scenario, we thus obtain a set of mitigated hours $M$ and compute the average impact of mitigation on the clearing price and on consumer surplus, respectively as~$\frac{1}{|M|}\sum_{t \in M} (P^*_t - P^m_t)$ and~$\frac{1}{|M|}\sum_{t \in M} D_t \cdot (P^*_t - P^m_t)$. 
\newline \newline 
Due to the limitations of publicly available data, our simulation makes some simplifying assumptions. First, the demand-supply equality is included with an inequality sign: therefore, the constraint accounts for the additional reserve demand specific to each load area, which are not reported in the load forecast. Second, to exclude the impact of congestion and other types of regulation on the historical nodal prices, we restrict our market-level simulation only to hours in 2019 that were neither mitigated nor congested (96.9\% of the year). Finally, we do not include start-up and grid-related unit constraints, as neither the grid location nor the fuel type of the units are available. As a consequence, the model constructs a market-level merit-order curve using solely real-time incremental bids. Although merit-order models disregard commitment decisions and related costs, this simplification is comparably accurate to depict the real-time market, as most commitment decisions are made in the day-ahead market~\cite{nyiso_manual}.  
\section{Results} \label{sec:results}
\subsection{Market-level response}
\begin{figure}
    \centering \includegraphics[width=0.9\textwidth]{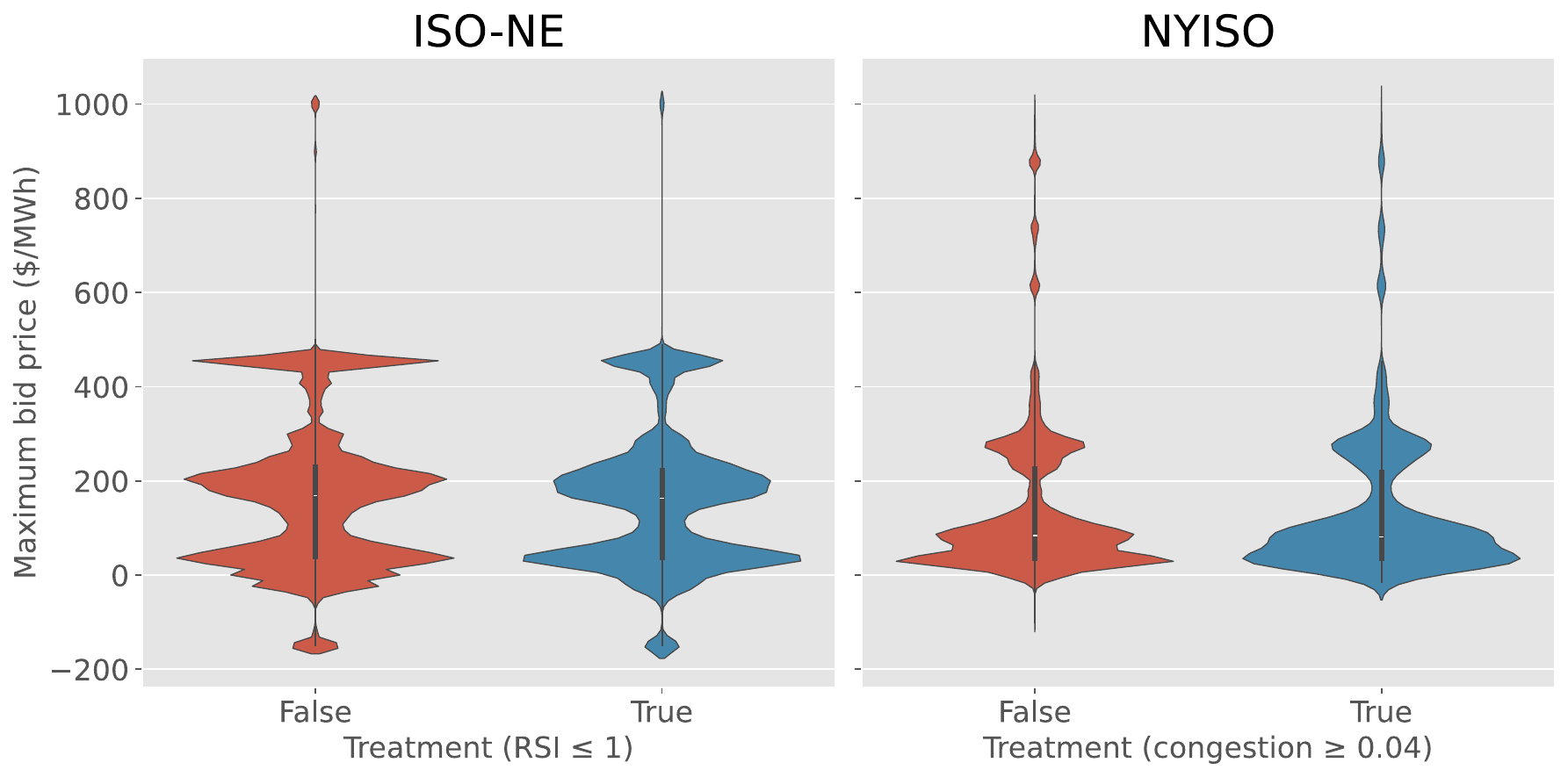}
    \caption{Distribution of maximum bid prices without and with treatment, i.e.,~active AMP screening in 2019.}
    \label{fig:bids_violinplot}
\end{figure}

In Table~\ref{tab:market_rdd} we present the results of the market-level regressions for the two markets, which include the main model specification -- a sharp RDD with a narrow bandwidth -- and two sensitivity tests, which use respectively a fuzzy RDD and a wider bandwidth. We are interested in the treatment coefficient, or LATE, which captures the causal effect of conduct screening activation (Step 2 of AMP) around the cutoff value for the structural variable (Step 1). This coefficient reflects how maximum bids change, on average, as a consequence of the discontinuous activation of a unit-specific conduct mitigation threshold once the cutoff of the structural test is exceeded. The pooled regressions for both markets indicate that the estimated local treatment effects align with a strategic response to AMP, which however is not statistically significant at a market level. In other words, while the negative coefficient suggests a reduction in the maximum bids of generation units when conduct screening becomes active, in none of the analyzed specifications we can reject the null hypothesis that this coefficient is, in fact, equal to zero. Hence, we cannot find evidence for a systematic, market-wide adjustment of bidding behavior as a response to the regulation, which would be consistent with a binding price cap and successful deterrence.
\newline \newline
LATE point estimates are similar in magnitude in both markets and negative, reflecting a downward adjustment in bid prices of around -1 to -2 \$/MWh. A negative treatment coefficient is consistent with a more careful behavior during screening, as bidders try to remain within their conduct tolerance thresholds in order not to trigger the penalty. LATE estimates in ISO-NE are below -1 \$/MWh for both the sharp and the fuzzy specification with a narrow bandwidth, while a wider bandwidth pushes the point estimate closer to zero. The identified LATE is moderately robust to the sensitivity tests presented in Table~\ref{tab:sensitivity}, i.e., a 50\% increase or decrease of the bandwidth for local estimation and the additional of a quadratic polynomial to the regression. In NYISO, treatment effects are below -1 \$/MWh when a narrow bandwidth is applied, and around -2 \$/MWh with a wider bandwidth. Additional sensitivity analyses confirm the main results, as none of these specifications provides sufficient evidence that the treatment coefficient differs from zero. 
\newline \newline 
The direction and magnitude of the structural coefficients in the regression are consistent with the adjustment strategy conceptualized in Figure~\ref{fig:example}. The estimated coefficients of the score variable are positive (main effect), while the coefficients for its interaction with treatment ($\tilde{S}\times T$) are negative. In other words, before the cutoff, bid prices increase with decreasing RSI or increasing congestion, which are associated with increased ability to exercise market power. Beyond the cutoff, coefficients flip to negative (combined main effect plus interaction effect), which corresponds to a decrease in bid prices as the relative score variable increases (i.e.,~as bid screening becomes increasingly likely). However, as visible from Figure~\ref{fig:bids_violinplot}, the overall distribution of maximum bid prices remains similar with and without active AMP screening.
\newline \newline
Reference levels represent a competitive price benchmark for the bidding unit and are particularly important in determining maximum bid prices, displaying positive and statistically significant coefficients in both markets. All else the same, it is reasonable to expect a positive relationship between competitive reference levels and maximum bids, as reference levels give an indication of overall marginal cost and average markup of a unit, as well as the leeway it has within conduct screening. Gas prices are also included as a control variable as gas power plants accounted for 50-60\% of the 2019 generation fleet in these two markets and, unlike in Europe, US gas plant operators tend to face direct exposure to short-term volatility in natural gas prices~\cite{alterman2012}. The magnitude and significance of gas prices are moderate in NYISO, but stronger in ISO-NE, which is consistent with the higher prices of natural gas on the New England gas pipeline~\cite{isoexternal2018}. Overall, the R$^2$ of the regressions indicates moderate explanatory power (0.50 -- 0.58), strengthening the validity of the model.   
\newline \newline 
Additional statistical analyses of maximum bids corroborate our findings. Overall, 87.71\% of the bids in ISO-NE are between 0 and 800 \$/MWh; lower bids usually signal must-run inflexibility, while higher bids signal the unavailability of (part of) the generation capacity~\cite{Reguant.2014}. Most generation units in this market display a low correlation of incremental bids with load forecasts, which could already indicate limited responsiveness to pivotality. In 2019, the median correlation between load forecasts and incremental bids in ISO-NE was less than 0.04: this is reflected in a correlation of around 0.40 between real-time clearing prices and load, which was lower than for European markets such as Spain or France (respectively, 0.53 and 0.74~\cite{entsoe_data}). Instead, we find a much stronger link to gas prices, with a median correlation between maximum bids and gas prices of around 0.20. The weak relationship between incremental bids and load levels can be partly attributed to the prevalence of bilateral contracts, which contribute to the large share of must-run and price-insensitive generation: as of 2024, price-insensitive generation still accounted for nearly 70\% of total generation in ISO-NE~\cite{isone2024}.

\begin{table}[htbp]
    \centering
    \begin{tabular}{p{3cm}*{6}{c}} 
   \toprule  & \multicolumn{3}{c}{\textbf{ISO-NE}} & \multicolumn{3}{c}{\textbf{NYISO}}\\
   \midrule 
   \textbf{Specification}             & Narrow          & Fuzzy           &  Wide           & Narrow         & Fuzzy       & Wide \\\midrule
    $T$                      &  -1.100        & -1.437          &-0.6582          & -0.9602       & -0.8018     & -2.073 \\%-0.0756 -1.153 
                             &(0.6591)        &(0.8241)         &(1.278)          & (2.025)       &(2.123)      & (1.058)\\%(0.6907)(2.425)
    $\tilde{S}$              & 19.01          & 19.53           &4.756            & 0.2277        &0.2242       & -0.5893$^{***}$\\%6.639 / 26.42 
                             & (12.11)        &(12.43)          &(3.494)          & (0.3570)      &(0.3558)     & (0.1462)\\%(9.078)  /(14.56)
    $\tilde{S} \times T$     & -67.27$^{***}$ &-65.41$^{**}$    &-51.55$^{***}$   & -0.9909       & -1.093      & 1.176$^{*}$ \\%-41.22$^{*}$ /-27.69
                             & (21.23)        &(20.47)          & (14.63)          &(1.988)       &(2.054)& (0.5566)\\%(18.30) /(23.19)  
    Ref.~level              & 0.5401$^{***}$ &0.5401$^{***}$   &0.5812$^{***}$   & 0.5941$^{***}$& 0.5941$^{***}$&0.5842$^{***}$\\%0.4942$^{**}$/0.6122$^{**}$
                             & (0.1452)       &(0.1452)         &(0.1320)         & (0.1729)      &(0.1719)     &(0.1719)\\%(0.1732)/(0.1826)
    Gas price                & 15.66$^{*}$    & 15.65$^{*}$     &7.756            & 2.596         &2.594        & 4.962\\%23.70$^{***}$/4.639
                             & (6.842)        &(6.844)          &(7.877)          & (4.246)       &(4.246)      & (4.913)\\%(6.659)(4.272)
    \midrule 
   \textbf{Tot.~observations}       & 729,904        & 729,904         &1,845,460        &474,170        &474,170      & 729,765\\
   \textbf{R$^2$}                    & 0.55443        &0.55443          & 0.58096          & 0.50700       &0.50700     & 0.50873 \\%Narrow: 0.53738 / 0.50269
   \textbf{Within R$^2$}              & 0.13131        &0.13131          & 0.13749          & 0.13722       &0.13722     & 0.13412\\%Narrow: 0.11473/ 0.14480
   \textbf{Fixed effects}            & \multicolumn{6}{c}{Bidder} \\
                             \bottomrule 
\multicolumn{7}{l}{{\footnotesize Significance codes: ***: 0.001, **: 0.01, *: 0.05. Standard errors (clustered by bidder)}}\\
\multicolumn{7}{l}{{\footnotesize in parentheses. Bandwidths ISO-NE: $\pm 0.2$ (narrow, fuzzy), $\pm0.5$ (wide). Bandwidths}}\\
\multicolumn{7}{l}{{\footnotesize NYISO: $\pm 3$~\$/MWh (narrow, fuzzy), $\pm20$~\$/MWh (wide).}}\\\\
\end{tabular}
\caption{Market-level regressions. Overall, adjustments to avoid conduct thresholds range between -1 and -2 \$/MWh in both markets.}
\label{tab:market_rdd}
\end{table}
%\subsection{Bidder-level response to AMP} \label{sec:bidder_regressions}
%%%[SUMMARY, BIDDER-LEVEL]

%From our results from the pooled regression, we can derive that point estimates exhibit the directions consistent with the expected strategic bidding behavior to avoid triggering the penalty. However, results are not significant on a market level. 

\subsection{Firm-level response}
\paragraph{ISO-NE}
The lack of statistical significance for the treatment in all model specifications hints at a weak market-level response to AMP. However, the consistently negative treatment effects suggest heterogeneity across bidders, i.e.,~some bidders may exhibit a stronger negative response to the regulation. To investigate this, we estimate the RDD model separately for each bidder; Table~\ref{tab:bidder_rdd} presents the results for the specifications previously introduced. Out of 81 bidders in the dataset, we exclude 7 bidders, which do not have enough data (e.g.,~because they were never pivotal) or enough variation in their maximum bids. We analyze around 35\% of the bids in the dataset, finding a significant firm response to the regulation for around 30\% of market bidders. This subset exhibits a lower median treatment effect of -4 to -5 \$/MWh compared to a median sample LATE of over -1\$/MWh. This is consistent with a stronger strategic adjustment of the bidding behavior of these bidders to avoid the penalty. In the wider model specification, the share of bidders with significant effects increases to almost 50\%, but the median treatment effect points to a more modest adjustment of over -3 \$/MWh.
\newline \newline
In ISO-NE terminology, a \textit{bidder} corresponds to a generation firm. Each firm is assigned an anonymized bidder ID; this enables comparisons between firms based on their number of assets and total fleet capacity. The median capacity of significant firms is around 243.5 MW, or 2.5 times larger than the median capacity of the sample (99.3 MW). This suggests that mitigation in New England is a strategic concern primarily for medium and large-scale firms. This finding is in line with the ISO's internal reports, which identify five large participants as responsible for 74\% of AMP activation in 2022~\cite{isone2022}.  In contrast, smaller firms rarely undergo a conduct-and-impact assessment and hence have little reason to adapt their bids to AMP.  This is a potential blind spot of the procedure, as large firms are not necessarily those bidding the highest prices in the market -- in fact, the median maximum bid prices of significant firms lies around 84.0 \$/MWh, below the sample median of 68.9\%. 
\newline \newline
\paragraph{NYISO}
We examined 75 out of 118 bidders in the NYISO market with sufficient historical data to reconstruct the reference levels. Overall, we find a median treatment effect of around -0.1 \$/MWh for the main model specification, and a null median treatment effect for the sensitivity test with wider bandwidths. However, also in this case, we identify a subset of bidders that respond to the regulation. This subset is smaller than in New England -- between 8 and 14\% of market bidders -- but exhibits a stronger treatment effect. For bidders with statistically significant treatment effects, the median downward adjustment associated to AMP activation lies between -8.6 and -9.9 \$/MWh. Similarly as in New England, the median installed capacity of these bidders is larger than the median capacity of bidders in the sample (respectively, 612.4 MW and 381.7 MW). 
\newline \newline
However, according to NYISO's jargon, a \textit{bidder} does not necessarily equate to a firm. While in ISO-NE, for 93.4\% of the units, bids are always placed by one bidder (its generation firm), in NYISO, the bids of a units can be associated to different bidder IDs: for some units, more than 15 different anonymized bidder IDs are used. In particular, accepted bids in the day-ahead market are assigned a real-time bid under NYISO's bidder ID, while if the bid is modified in the real-time market, the assigned bidder ID might be different, matching the party who made the change~\cite{nyisobiddata}. This introduces some selection bias, as the bidder ID depends on the acceptance status of the bid. Moreover, the anonymized data do not allow location matching, hence the analysis is diluted by bidders located in the New York City area: since this area always underlies screening for congestion, no adaptation to the activation of AMP can be observed there.
\newline \newline
Finally, while our analysis shows that a response to congestion exists for some of the bidders in this market, the results suffer from a number of limitations due to the lack of grid-level information. In particular, our structural test uses a lagged load-averaged congestion variable, whereas in congestion-based AMP, conduct screening is applied individually to each load area. This induces some measurement error which has a temporal component (because the indicator is lagged) and a spatial component (because the indicator is an aggregated market-level index). We conduct additional market-level regressions to assess the sensitivity of the results to the choice of the proxy score variable, reported in Table~\ref{tab:sensitivity}. We re-estimate the model using current shadow price averages with no lag, which we avoided in the main specification due to simultaneity between real-time prices and bids, and lagged maximum congestion across load areas (excluding New York City), which we avoided due to the lack of spatial consistency of the indicator. Finally, we test adding a second-order term to the main model specification. We find that results in NYISO are not robust to these changes: treatment coefficients remain non-significant and within the magnitude of 1 \$/MWh, but flip signs from negative to positive in some specifications.
\begin{table}
\centering
\begin{tabular}{p{2.8cm}*{6}{c}}\\
 \toprule  & \multicolumn{3}{c}{\textbf{ISO-NE}} & \multicolumn{3}{c}{\textbf{NYISO}} \\\midrule
 \textbf{Specification} & Narrow & Fuzzy & Wide & Narrow & Fuzzy & Wide \\\midrule
\textbf{Analyzed} \\
No.~of bidders  &\multicolumn{3}{c}{74} &\multicolumn{3}{c}{75}\\
(\% of total)   &\multicolumn{3}{c}{(91.4\%)} &\multicolumn{3}{c}{(63.6\%)}\\
No.~of bids  & 728,090 & 728,090  & 1,838,506  & 472,897 & 472,897  & 727,755  \\
(\% of total) & (34.7\%) & (34.7\%) & (87.6\%) & (61.9\%) & (61.9\%) & (95.3\%) \\
Median treat.~effect & -0.640 & -0.898 & -1.121 & -0.105 & -0.439 & 0.000 \\
(IQR) & (-4.2, 0.4)& (-4.7, 0.4) &(-3.4, 0.0) &(-6.6, 1.1) &(-8.1, 1.0) &	(-5.3, 2.3) \\
\textbf{Significant} \\
No.~of bidders & 26 &24  &40 & 17  & 10 & 15 \\
 (\% of analyzed) & (35.1\%) & (32.4\%) & (54.1\%) & (22.7\%)& (13.3\%) & (20.0\%)\\
No.~of bids  & 289,642  & 275,730  & 1,173,042   &124,057  & 128,772  & 201,570 \\
(\% of analyzed) & (39.8\%) & (37.9\%) & (63.8\%) & (26.2\%) & (27.2\%) & (27.7\%) \\
Median treat.~effect & -4.386 & -5.032 & -2.616 & -8.636  & -9.874  & -4.575 \\
 (IQR) & (-5.0, -1.2) & (-7.0, -1.5) & (-4.0, 2.0) & (-13.0, 8.6) & (-13.7, 3.1) &	(-9.7, 7.8) \\
\midrule
\textbf{Tot.~bids}&\multicolumn{3}{c}{2,099,254}&\multicolumn{3}{c}{764,030}\\
\textbf{Tot.~bidders}&\multicolumn{3}{c}{81} &\multicolumn{3}{c}{118} \\
\textbf{Control variables}&\multicolumn{6}{c}{Ref.~level, gas price}\\
\bottomrule
\multicolumn{7}{l}{{\footnotesize A bidder and its related bids are classified as significant when the $p$-value of its treat.}}\\
\multicolumn{7}{l}{{\footnotesize coefficient is 0.05 or lower. Bandwidths ISO-NE: $\pm 0.2$ (narrow, fuzzy), $\pm0.5$ (wide).}}\\
\multicolumn{7}{l}{{\footnotesize Bandwidths NYISO: $\pm 3$~\$/MWh (narrow, fuzzy), $\pm20$~\$/MWh (wide).}} \\\\
\end{tabular}
\caption{Summary of the bidder-level regressions. The median LATE of significant bidders is 2 to 10 times lower than in the analyzed sample.}
\label{tab:bidder_rdd}
\end{table}
\subsection{Welfare implications of tightened regulation}
\begin{figure}[htbp]
    \centering
    \includegraphics[width=0.9\textwidth]{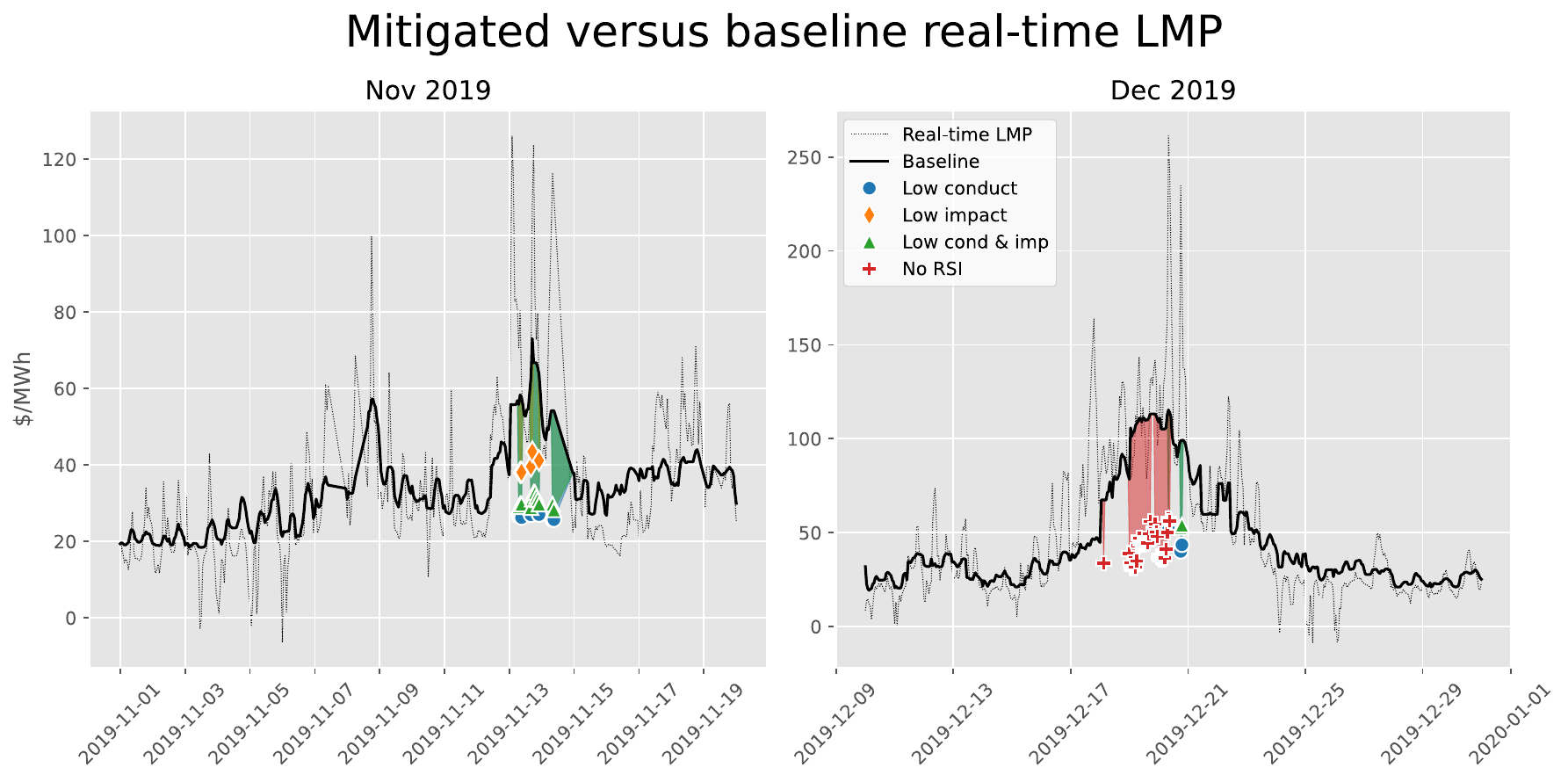}
    \caption{Real-time price in Nov-Dec 2019 for the baseline scenario and simulations with tighter AMP thresholds. ISO-NE's historical hub LMP is shown for reference (dotted).} 
    \label{fig:simulated_prices}
\end{figure}

One of the main takeaways from the results of the RDD is that the conduct thresholds may not be sufficiently stringent to be binding and therefore do not induce response in the electricity prices set by bidders. We draw on this finding and simulate real-time prices under stricter AMP regulation, building a merit-order model for the ISO-NE market in 2019. We exclude from the simulation hours that were either mitigated by other types of regulation or congested in real-time. Table~\ref{tab:simulated_prices} summarizes the results, where the first panel represents a scenario where AMP under current thresholds is implemented, which is used as a baseline for the analysis~\footnote{We did not use the historic real-time price as a baseline, as scenario outcomes might deviate from true prices for other reasons than AMP mitigation.}. Despite the simplifying assumptions imposed by the unavailability of location and fuel data, real-time prices from the baseline scenario closely approximate the real-time LMP at ISO-NE's internal hub, with a median price deviation of 1.01 \$/MWh and a median absolute error of 4.82 \$/MWh.
\newline \newline
The baseline scenario yields no mitigated hours: although the pivotality test fails for one or more suppliers in 16.3\% of the 8,493 simulation hours (Fig.~\ref{fig:amp_flowchart}, Step 1), and the conduct screening subsequently fails for 9.71\% of the hours (Step 2), in none of these cases leads to a price impact that exceeds the threshold (Step 3). This is in line with the ISO's historical statistics on AMP, which also reported no mitigation for 2019~\cite{isone2018}. In the other scenarios, expanding the scope of bids tested for mitigation (stricter Step 1 and 2) proves more effective than tightening the thresholds for price impact (Step 3). Removing the pivotality test -- thereby applying the conduct-and-impact test to all market participants -- has the greatest effect, leading to more than 30 mitigated hours. Lowering conduct thresholds results in three times more mitigated hours with respect to lowering impact thresholds (respectively, 15 and 5 mitigated hours). During mitigated hours, prices fall on average by over 64 \$/MWh when removing the pivotality test, and by over 35 \$/MWh when reducing the conduct thresholds instead. Despite more stringent tests and occasional market interventions, the average real-time price in all simulations remains around 31 \$/MWh. The findings indicate that enhancing AMP could generate an additional \$350,000 to \$980,000 in buyer surplus for each hour of mitigation achieved. In total, buyer surplus could increase by over \$9 million under tighter conduct thresholds, and even exceed \$30 million if the pivotality test is removed. 
\newline \newline
Tightening AMP thresholds evokes two concerns: first, that stringent AMP may lead to excessive market intervention, constantly disturbing the auction mechanism; second, partly following the first, that very stringent AMP may prevent scarcity pricing and cost recovery (e.g.~fixed cost, start-up cost etc.) -- up to the point of negative producer surplus if reference levels are below marginal cost. Regarding the first point of concern, we find that our implemented changes do not trigger constant market intervention. In fact, as shown in Figure~\ref{fig:simulated_prices}, mitigated hours (less than 2 days in total) are limited to a 2-month period (November-December 2019), during which high load and reduced generation availability resulted in an unusually high number of pivotal suppliers. In particular, 16.5\% of the real-time bidding capacity was unavailable in November, compared to a yearly average of 8.7\%. Regarding the second point of concern, it should be clarified that, despite the comparably low available capacity, the real-time price spikes in this period cannot be attributed to scarcity pricing, as ISO-NE has a separate protocol that allows prices to exceed market caps in case of supply shortage ~\cite{isone2018}. Moreover, several other aspects of the market design provide opportunities for cost recovery. First, both ISO-NE and NYISO run capacity markets for investment cost recovery. Second, conduct thresholds are set higher than reference levels, which leaves a tolerance buffer to break even and some fixed cost recovery. Third, the AMP procedure is applied only to incremental bids, so that additional cost recovery via complex bids is not impacted. Lastly,~\cite{adelowo2024} perform a welfare analysis on simulated greenfield AMP in the Iberian data and find that AMP deliver positive producer surplus, which can even be increased by imporving the precision of reference levels. 
\begin{table}[htbp]
    \centering
   \begin{tabular}{p{2.8cm}*{5}{c}}
 \toprule & \multicolumn{5}{c}{\textbf{ISO-NE}} \\\midrule
 \textbf{Scenario}& Baseline & Lower & Lower & Lower cond. & No pivot. \\
 & & conduct & impact &  \& impact & test\\\midrule
 \textbf{Structural} & RSI $\leq 1$ & RSI $\leq 1$ & RSI $\leq 1$ & RSI $\leq 1$ & - \\
 \textbf{cutoff} \\
 \textbf{Conduct} & 100 \$/MWh  & 50 \$/MWh  & as in  &  75 \$/MWh  & as in  \\
 \textbf{threshold} & or 300\% & or 150\% & base & or 200\% & base \\
 \textbf{Impact} & 100 \$/MWh  & as in  &  50 \$/MWh  & 90 \$/MWh  & as in \\
 \textbf{threshold} & or 200\% & base & or 150\% & or 175\% & base \\\midrule
\textbf{Mitigated} & 0 & 15 & 5 & 15 & 32\\
 \textbf{hours} \\
\textbf{Avg.~clearing} & 31.35 &  31.26 & 31.33 & 31.27 & 31.07 \\ % load-weighted price average of all hours where 
\textbf{price (\$/MWh)} \\
%\textbf{Weighted avg. price (\$/MWh)} & 31.347 &  31.262 & 31.332 & 31.271 & 31.068 \\ % load-weighted price average of all hours where 
\textbf{Avg.~decrease in}  & - & 39.35 & 23.06 & 35.37 & 64.68\\ % simple price 
\textbf{price (\$/MWh)} \\
%\textbf{Avg. mitigation \newline impact (\$/MWh)}  & - & -39.353 & -23.056 & - 35.367& -64.684\\ % simple price average of all hours where price != benchmark price
\textbf{Buyer surplus \newline increase  (\$)} & - &9,606,561 & 1,747,825 & 8,610,908 & 31,398,326\\
%\textbf{Hourly buyer surplus (\$/h)} & - &640,437.4 & 349,565 & 574,060.5& 981,197.6\\
 \bottomrule \\
    \end{tabular}
    \caption{Results of ISO-NE real-time price simulations for 2019. The baseline scenario implements the current AMP and is used as benchmark.}
    \label{tab:simulated_prices}
\end{table}

\section{Discussion} \label{sec:discussion}
While our estimated treatment effects are generally negative, for the majority of firms we do not find a negative adjustment of bid prices to avoid the price-cap-and-penalty regulation imposed by AMP. The heterogeneous response between firms is insufficient to detect a consistent, market-wide impact of AMP regulation on bid prices. Our results may have two main implications, both of which point towards inefficiency of the current regulatory framework. It may be that, even during supply scarcity, the analyzed markets do not experience market power abuse, and hence AMP is superfluous; or that market power abuse exists but the dynamic, temporary price thresholds from the AMP are loose enough to allow this during active screening. This would imply that the regulation is ineffective in reaching its deterrence goal and, therefore, inefficient. The former option seems unlikely, given that other studies find pricing patterns consistent with Cournot-style market power exertion in times of supply scarcity~\cite{fladung2024electricity}. The latter option describes a more plausible scenario, in particular given that the mitigation of reliability units~\footnote{Here, reliability generation refers to redundant capacity which is committed due to operational requirements.}, which is less predictable and uses more stringent thresholds, strikes considerably more often~\cite{isone2019}. 
\newline \newline
In all cases, our findings provide evidence that the lack of AMP mitigation is \textit{not} a consequence of successful deterrence, as this should have led to significant discontinuities in bidding behavior around the screening cutoffs. The notion from~\cite{goldman2004review} should therefore be carefully interpreted: effective AMP exhibit low trigger rates, however, rare triggering is not at all a sure sign of effective AMP. This further underscores the importance of carefully choosing price caps and motivates a simulation-based selection of AMP thresholds. With our simulations, we show that a better calibration of AMP can increase the effectiveness of this regulation. At the same time, tightening the conduct thresholds or eliminating the pivotality requirement does not automatically lead to a substantial interference with market operations (32 mitigated hours per year or less). The buyer surplus of improving AMP could be as high as \$31 million dollars per year, although outcome deviations are likely, as market participants respond to regulation. To accurately estimate welfare benefits, future research could include learning and strategic agents in the simulation. \newline \newline
\paragraph{Limitations}
One of the key limitations of the pooled method concerns the fact that it only estimates \textit{average} behavior. Not finding any systematic market-level differences in maximum bids around the structural threshold, does not imply that AMP is entirely disregarded by firms. This is supported by the significant LATE results for a subset of firms in our firm-level analysis. Alternatively, some firms may simply be unable to accurately anticipate their screening status, resulting in no meaningful regulatory adjustment. The heterogeneous response of these bidders -- which notably tend to be smaller, in line with previous empirical findings~\cite{hortacsu2005understanding} - may thus weakening the pooled market-level estimates. In fact, in simulations where rational bidders are informed of screening activation, they have been shown to bid up to the mitigation threshold~\cite{entriken2005agent}. It is therefore possible that strategic behavior will occur during one-off, extreme situations where conduct-and-impact screening is certainly activated (e.g.,~extremely high load or congestion). 
\newline \newline
Finally, the scope of the available data limits the precision of our analysis. Reference levels are confidential to the ISO and the respective generation firm and may be computed using three different procedures. Instead, we computed reference levels by approximating the offer-based procedure using historical incremental bids. Similarly, the structural test used in this study are not identical to those applied by the ISOs. For ISO-NE, we derive a Pivotal Supplier Test based on the available open source data, which may not coincide with the full information at ISO’s disposal. In NYISO, limited data availability -- specifically, missing information on unit location -- means that our analysis must be restricted to market-level congestion.
\section{Conclusion} \label{sec:conclusion}
Market power is a widespread problem in electricity markets and its mitigation through ex-post procedures is lengthy and expensive. Therefore, many US markets apply ex-ante mitigation in their electricity wholesale auction markets, in the form of automated mitigation procedures (AMP). To the best of our knowledge, this study provides the first causal assessment of firms' bidding responses to ex-ante market power mitigation in wholesale electricity markets, focusing on hourly unit-level bid prices from real-time auctions in New England and the State of New York in 2019. Currently, AMP are the only regulation whose goal is to preemptively limit market power abuse of generation firms. They are multi-step price-cap-and-penalty procedures: when structural market conditions favor market power abuse, a conduct assessment of the bids for electricity generation is activated. Bid prices are thus temporarily capped to conduct thresholds and, if market-distorting behavior is found, mitigated down to a competitive reference level. As this effectively constitutes a penalty, AMP activation should incentivize firms not to exceed the temporary price caps. We investigate the deterrent power of AMP by testing if firms reduce their bid prices to avoid the penalty, as soon as the conduct assessment becomes active. To causally identify this effect, we employ a regression discontinuity design, which exploits the discontinuous activation of conduct assessment once a structural market index is exceeded.
\newline \newline
In the analyzed markets and under the current AMP thresholds, the estimated market-level impact of the regulation on the bids submitted by generation units is a 1-2 \$/MWh decrease in price. However, this impact is not statistically significant; in other words, the applied mitigation thresholds, defined in the early 2000s, do not appear to have a market-wide deterrent effect. Individual analyses reveal heterogeneity in bidder behavior and find a subset of bidders with a statistically significant effect. These bidders adjust their bid prices in response to AMP regulation: at the median, they reduce them by 4-5~\$/MWh (ISO-NE) and 9-10~\$/MWh as a response to active AMP regulation. In line with previous empirical findings~\cite{hortacsu2005understanding}, we find that larger firms are likely to better predict the activation of the regulation and be more responsive.
\newline \newline
Given the importance of setting appropriate tolerance thresholds for the electricity bids of generation units, we construct a simulation of the real-time market in New England and showcase the benefits of empirically calibrating mitigation thresholds. Depending on the scenario, we obtain buyer surplus gains between 350 and 980 thousand dollars per additional mitigated hour. We also show that, despite concerns of overmitigation associated with tighter mitigation thresholds, the most stringent AMP in our simulation would not result in more than 32 mitigated hours per year. However, more precise estimates would require more data, and in particular more transparency on the reference level of the units and on their location. In general, the results underscore both the potential and limitations of automated mitigation as a tool to preserve competitiveness in electricity markets and can serve as a blueprint for regulators to evaluate preventive measures against market power abuse. 
\section{Acknowledgements}
The project was funded by the Federal Ministry of Research, Technology and Space (BMFTR, project code: 16DKWN102) and is part of the German Recovery and Resilience Plan (DARP), financed by NextGenerationEU, the European Union's Recovery and Resilience Facility (ARF). J.A.~further acknowledges funding from the Extended Partnership Program “Network 4 Energy Sustainable Transition” - Acronym NEST, Program Code PE\_000021, CUP E13C22001890001, Notice No.~341 of  \newline 15/03/2022 - Piano Nazionale di Ripresa e Resilienza (PNRR), Mission 4 Istruzione e ricerca – Component 2 Dalla ricerca all’impresa – Investiment 1.3, funded by the European Union - NextGenerationEU. C.F.B.~further acknowledges travel funds from the German Energy Agency (DENA). \\\\ We thank Jacob Grindal, Paul Irvine (ISO-NE), Pradip Kumar (NYISO), Pallas Lee VanSchaick (Potomac Economics), Alice Lixuan Xu, Jorge Sánchez Canales and Lion Hirth (Hertie School) for their valuable insights. 
\section{Author contributions: CRediT}
C.F.B.: Conceptualization, Data curation, Formal analysis, Investigation, Methodology, Visualization, Writing - original draft, Writing - review \& editing.
J.A.: Conceptualization, Investigation, Methodology, Visualization, Writing - original draft, Writing - review \& editing.
P.D.: Supervision, Writing - review \& editing.
L.K.: Supervision, Writing - review \& editing.

\clearpage
\appendix

\section{Key market characteristics}
\setcounter{table}{0}
\renewcommand{\thetable}{A\arabic{table}}
\begin{table}[htbp]
    \centering
    \begin{tabular}{p{6cm}p{2cm}p{2cm}} \toprule
    & \textbf{ISO-NE} & \textbf{NYISO} \\\midrule
    \textbf{Installed capacity (GW)}   & &  \\%sources [NYISO]: 2019 Load & Capacity Data Report (Table III-3b)  [ISO-NE]: ISO New England 2019 Regional System Plan (Figure 7-2)
    {Total} & 33.43 & 41.79  \\ 
     {Gas-fired}   & 16.54 & 24.64  \\%[NYISO]: consider also oil & gas units
    \textbf{Average price (\$/MWh)} & &  \\
     {Day-ahead} & 32.97 & 27.98 \\ 
     {Real-time}& 32.23 & 27.54 \\
    \textbf{Constrained hours (real-time)} & & \\
     {Congestion} & 1,280 & 2,504 \\
     {Pivotality} & 1,486 & - \\\bottomrule
     \multicolumn{3}{l}{\footnotesize  Pivotal hours omitted in NYISO due to incomplete firm data.}\\\\
    \end{tabular} %Gas-fired installed capacity includes fuel-switching (oil and gas) sources. Prices are load-weighted averages. Congestion: at least one zone has shadow prices of 0.04 \$/MWh and above. Pivotality: at least one company is pivotal.
    \caption{Analyzed markets in 2019. Price averages refer to the hub LMP for ISO-NE and the load-weighted zonal LMP in NYISO.} 
\label{tab:market_characteristics}
\end{table}
\section{Data sources}
\setcounter{table}{0}
\renewcommand{\thetable}{B\arabic{table}}
\begin{table}[htbp]
    \centering
    \begin{tabular}{ll} \toprule
    \textbf{Data} \\
     Gas prices &  {\footnotesize\href{https://www.investing.com/commodities/natural-gas-historical-data}{investing.com/commodities/natural-gas-historical-data}} \\\midrule
    \textbf{ISO-NE} & {\footnotesize \href{https://www.iso-ne.com/isoexpress/web/reports}{iso-ne.com/isoexpress/web/reports}} \\
       Unit bids & {\footnotesize\href{https://www.iso-ne.com/isoexpress/web/reports/pricing/-/tree/real-time-energy-offer-data}{pricing/-/tree/real-time-energy-offer-data}} \\
      Prices & {\footnotesize\href{https://www.iso-ne.com/isoexpress/web/reports/pricing/-/tree/lmp-by-node}{pricing/-/tree/lmp-by-node}} \\
      Tot.~load & {\footnotesize\href{https://www.iso-ne.com/isoexpress/web/reports/load-and-demand/-/tree/dmnd-da-hourly-cleared}{load-and-demand/-/tree/dmnd-da-hourly-cleared}} \\
      Load forecast & {\footnotesize\href{https://www.iso-ne.com/isoexpress/web/reports/load-and-demand/-/tree/three-day-reliability-region-demand-forecast}{load-and-demand/-/tree/three-day-reliability-region-demand-forecast}} \\
      Tot.~reserves & {\footnotesize\href{https://www.iso-ne.com/isoexpress/web/reports/grid/-/tree/ancillary-hourly-rr}{grid/-/tree/ancillary-hourly-rr}} \\\midrule
    \textbf{NYISO} & {\footnotesize\href{https://mis.nyiso.com/public}{mis.nyiso.com/public}} \\
     Unit bids & {\footnotesize\href{https://mis.nyiso.com/public/P-27list.htm}{P-27list.htm}} \\
     Load forecast & {\footnotesize\href{https://mis.nyiso.com/public/P-7list.htm}{P-7list.htm}}  \\
     Prices & {\footnotesize\href{https://mis.nyiso.com/public/P-4Alist.htm}{P-4Alist.htm}} 
    \\\bottomrule\\
    \end{tabular} 
        \caption{Link to data sources. For ISO data, we provide the directory path and related subdirectories.}
\label{tab:sources}
\end{table}
\clearpage
\section{Example of reference level}
\setcounter{figure}{0}
\renewcommand{\thefigure}{C\arabic{figure}}
\begin{figure}[htbp]
\centering    \includegraphics[width=0.9\textwidth]{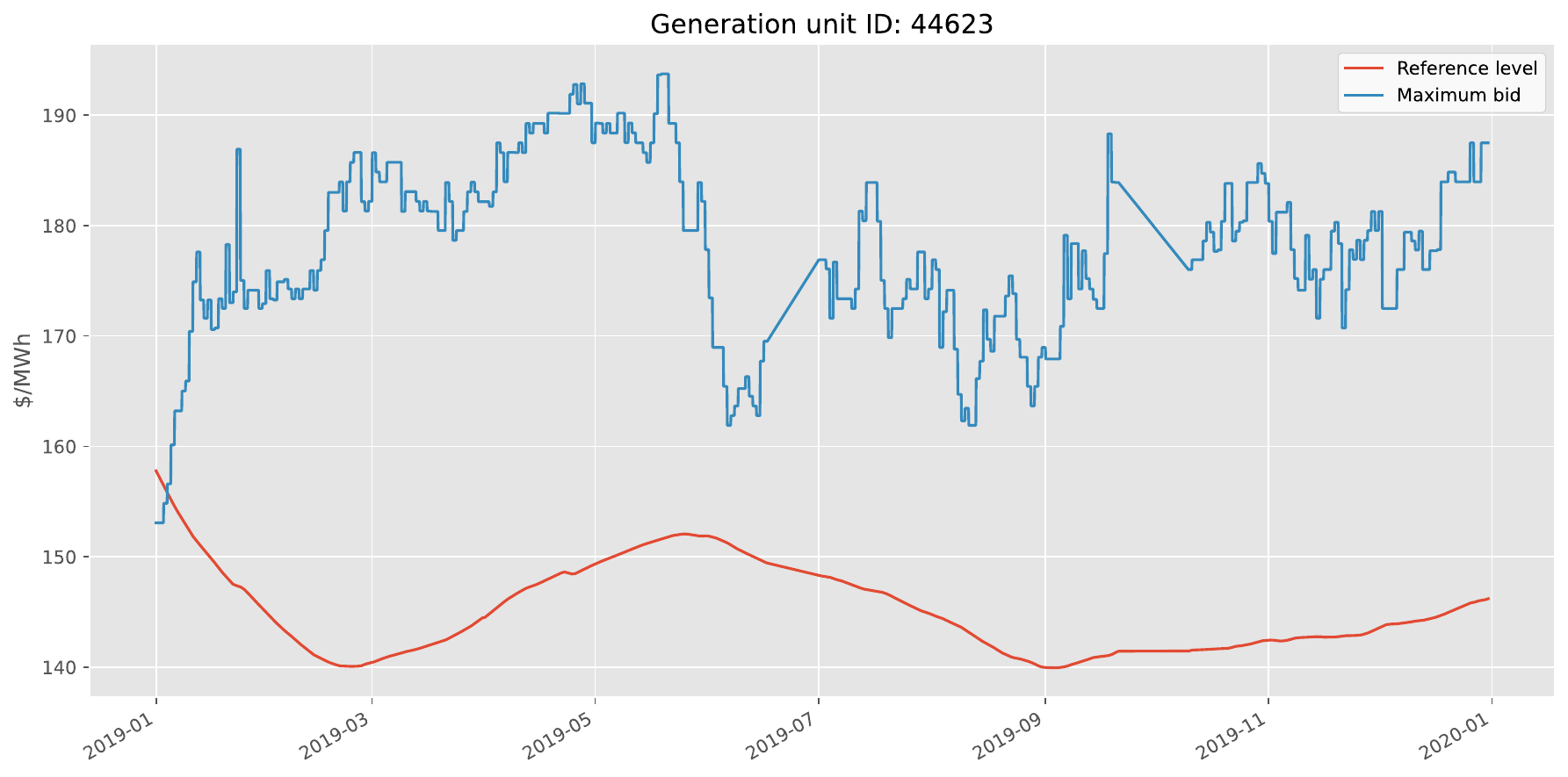}
    \caption{Reference level and maximum incremental bid for a generation unit in ISO-NE. The reference level serves as a benchmark for the conduct-and-impact assessment of unit bids.}
    \label{fig:ref_level}
\end{figure}
\section{Bandwidth selection}\label{sec:bandwidth}
\setcounter{figure}{0}
\renewcommand{\thefigure}{D\arabic{figure}}
\setcounter{table}{0}
\renewcommand{\thetable}{D\arabic{table}}
In RDD, fitting (high-order) polynomials to the entire dataset can lead to noisy and treatment effect estimates, which are highly sensitive to the assumed polynomial parametrization~\cite{huntington2021effect,gelman2019high}. In the selected cases, the score variable is not smoothly distributed (see Table~\ref{tab:score} and Figure~\ref{fig:score}), and fitting the regression to the whole dataset would place substantial weight on observations far from the cutoff. This may lead to an inappropriate regression fit around the cut-off, which may bias our results. In the case of ISO-NE, the distribution of the score variable (RSI) varies between bidders: at a market level, a pivotal supplier was present in only 17\% of the hours in 2019, but the number of hours in which a bidder was pivotal is even lower for smaller firms. Therefore, less than a quarter of the observations fall close to the cutoff value of 1. In the case of NYISO, the score variable (load-averaged shadow price) is highly skewed: in other words, the market is mostly not congested, but can experience extreme situations where the average shadow price drops to -600 \$/MWh or rises above 70 \$/MWh. 
\newline \newline
Following established literature, we adopt a locally linear approach, selecting observations located within a symmetric bandwidth around the cutoff value. To determine an appropriate bandwidth, we construct index duration curves around the cutoff, choosing the percentage of the dataset to retain. In the main specification, we select a narrow bandwidth of $\pm 0.2$ around the cutoff of 1 for ISO-NE and of $\pm 3$ around the cutoff value of 0.04 \$/MWh for NYISO, which corresponds to approximately 30\% and 60\% of the data. An alternative, wide specification is provided with bandwidths of 0.5 for ISO-NE and 20 \$/MWh in NYISO. Additional sensitivity tests on the selected bandwidth and fitted polynomial are presented in Table~\ref{tab:sensitivity}. 
\clearpage
\begin{figure}
    \centering
    \includegraphics[width=0.9\textwidth]{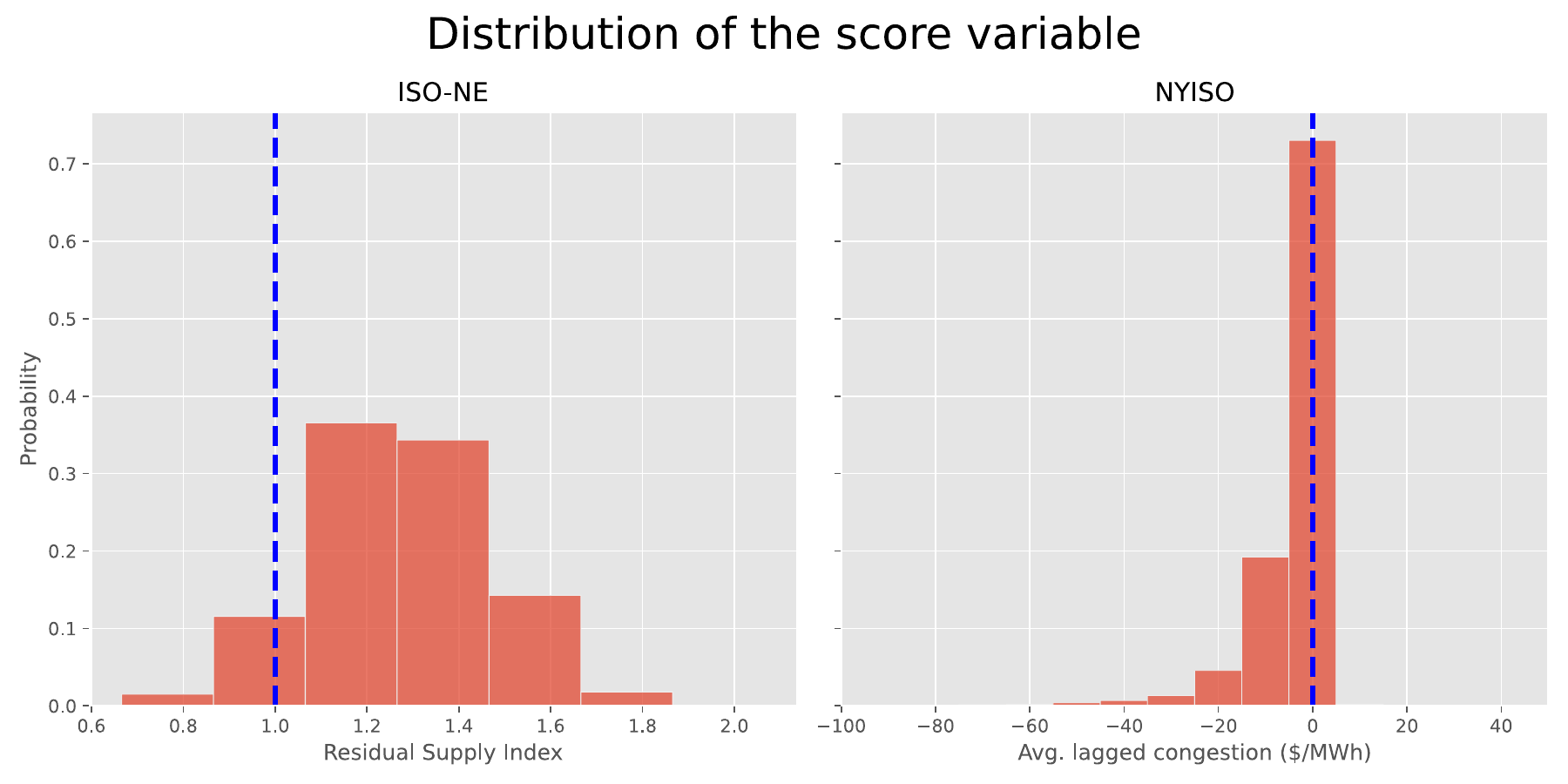}
    \caption{Distribution of the score variable.}
    \label{fig:score}
\end{figure}
\begin{table}[htbp]
\centering
    \begin{tabular}{l*{7}{p{1.35cm}}}\toprule
    &\textbf{Cutoff value} & \textbf{Min} & \textbf{1$^{\text{st}}$ quartile} & \textbf{Median} &\textbf{Mean} & \textbf{3$^{\text{st}}$ quartile} & \textbf{Max} \\\midrule
         \textbf{ISO-NE}& 1 & 0.6658 &  1.1485  & 1.2677 &  1.2740 & 1.4034 &  1.8796 \\
         \textbf{NYISO} & 0.04 &-615.04 & -5.4722 & -1.3382 &  -4.7039 &   -0.0014 &   73.6538 \\\bottomrule\\
    \end{tabular} 
    \caption{Descriptive statistics of the score variables.}
    \label{tab:score}
\end{table}
\section{Fuzzy regression discontinuity}\label{sec:fuzzy} 
\setcounter{figure}{0}
\renewcommand{\thefigure}{E\arabic{figure}}
If our estimate of the score (running) variable, $\hat{S}$, is only an approximation of the true one, $\tilde{S}$, or there is some uncertainty around the treatment assignment (e.g.,~because bidders imperfectly predict when they will be treated), there might be some noise in the treatment around the cutoff. 
We address this uncertainty by introducing a fuzzy RDD. Without loss of generality, we assume that $\tilde{S}$, the true score variable, is centered with a cutoff $c := 0$, and the received treatment is $T := \mathbb{1}\{\tilde{S}\geq 0\}$. We assume that our estimator $\hat{S}$ is asymptotically correct, but we acknowledge some randomly distributed measurement noise due to variance in the data, i.e.,~$\hat{S} = \tilde{S} + \varepsilon$ with $\varepsilon \sim \mathcal{N}(0, \sigma^2)$. We can therefore write the probability of treatment $p$ as:
$$p = P(\tilde{S} \geq 0) = P(\hat{S} - \varepsilon \geq 0) = P(\varepsilon \leq  \hat{S}) = \Phi\Big(\frac{\hat{S}}{\sigma}\Big) $$
and use it to estimate a fuzzy RDD with continuous treatment in the immediate proximity of the cutoff, reflecting the uncertainty related to noise. Figure~\ref{fig:fuzzy_treatment} shows an example of a continuous fuzzy treatment for a score function between -1 and 1. We tested different implementations of the continuous treatment variable with standard variances of 0.01, 0.05 and 0.1; and reported values for $\sigma = 0.01$ in the results.
\begin{figure}[htbp]
    \centering
    \includegraphics[width=0.9\textwidth]{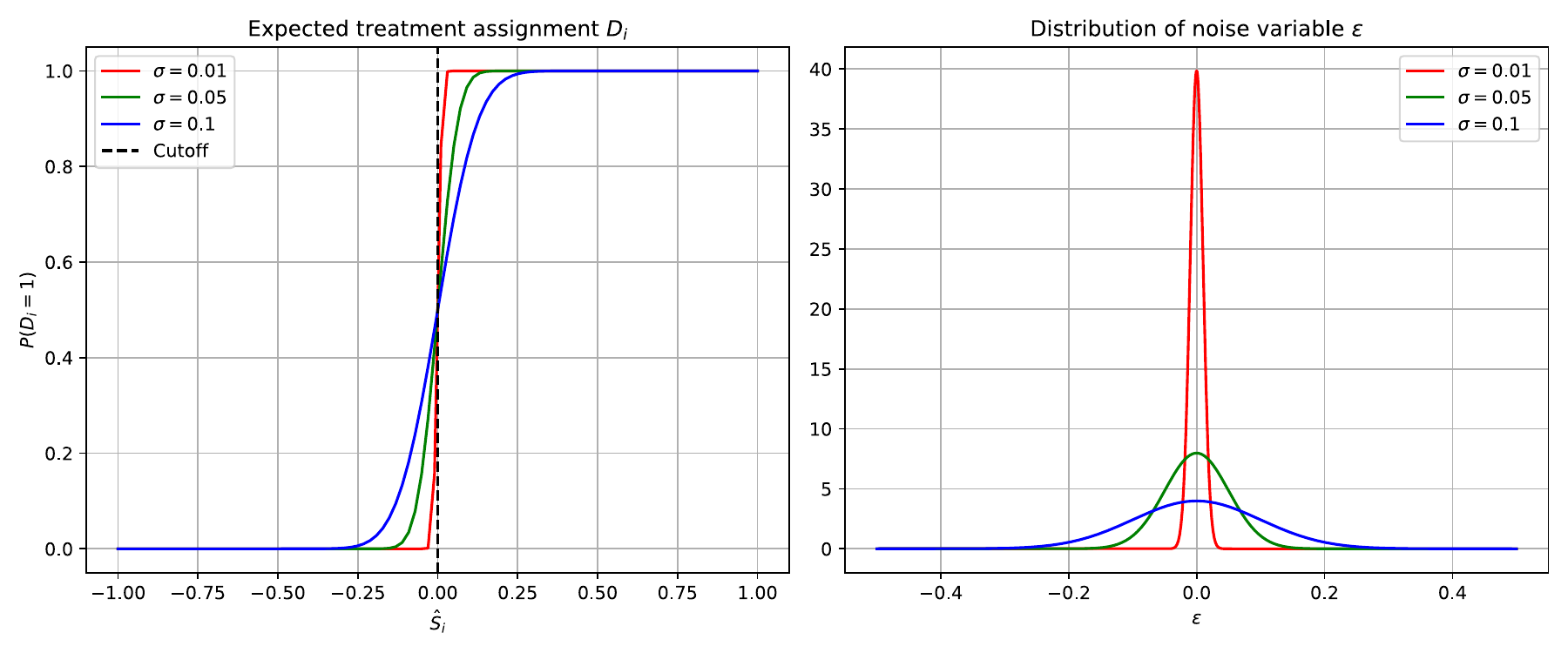}
    \caption{Probability of treatment assignment where $\hat{S}$ is the computed structural score, and $\varepsilon$ the random noise.}
    \label{fig:fuzzy_treatment}
\end{figure}

\clearpage
\section{Sensitivity tests}\label{sec:sensitivity}
\setcounter{table}{0}
\renewcommand{\thetable}{F\arabic{table}}
\begin{table}[htbp]
    \centering
\begin{tabular}{p{2.5cm}*{6}{c}}
\toprule  & \multicolumn{3}{c}{\textbf{ISO-NE}} & \multicolumn{3}{c}{\textbf{NYISO}}\\
   \midrule
    \textbf{Specification}             & Extra narrow          & Medium        &  Polynomial           & No lag         & Max cong.       & Polynomial \\\midrule
   %\emph{Coefficients}\\
    $T$                      &  -0.0756        & -1.306         &-0.2681          & 0.2634      & -0.3242     & 0.6193 \\%-0.0756 -1.153 
                             &(0.6907)         &(0.7752)         &(0.8623)          & (0.4083)       & (1.831)     & (3.453)\\%(0.6907)(2.425)
    $\tilde{S}$              & 6.639          & 12.61           & -9.324            & -0.4365$^{***}$        & -0.1380       & -0.5525\\%6.639 / 26.42 
                             & (9.078)        & (7.630)         & (12.71)        & (0.3570)      &(0.3558)     & (0.1462)\\%(9.078)  /(14.56)
$\tilde{S}^2$              & -          & -           & -131.4            & -        &-       & \\%6.639 / 26.42 
                             &        &          & (82.73)        & (0.3570)      &(0.3558)     & (0.1462)\\%(9.078)  /(14.56)
    $\tilde{S} \times T$     & -41.22$^{*}$ &-58.34$^{**}$    &-28.27  & -2.097$^{***}$       & -0.0357      & -6.029 \\%-41.22$^{*}$ /-27.69
                             &(18.30)        &(17.55)          & (25.80)         &(0.5702)       &(0.2253)& (7.700)\\%(18.30) /(23.19) 
        $\tilde{S}^2 \times T$              & -         & -           & 69.96            & -        &-       & 2.930\\%6.639 / 26.42 
                             &     &  & (161.7)       &      &    & (3.545)\\%(9.078)  /(14.56)
    Ref.~level          & 0.4942$^{**}$ & 0.5640$^{***}$   &0.5401$^{***}$   & 0.5942$^{***}$& 0.5940$^{***}$&0.5940$^{***}$\\%0.4942$^{**}$/0.6122$^{**}$
                             & (0.1732)      & (0.1346)        & (0.1452)        & (0.1729)      &(0.1729)     &(0.1729)\\%(0.1732)/(0.1826)
    Gas price                & 23.70$^{***}$    & 11.08    & 15.63$^*$            & 2.521         &2.566        & 2.597\\%23.70$^{***}$/4.639
                             & (6.659)       & (7.372)         & (6.854)         & (4.242)       &(4.213)      & (4.244)\\%(6.659)(4.272)
    \midrule 
   \textbf{Tot.~\newline observations}       & 314,591        & 1,189,516         &729,904       &474,170        &474,170      & 474,170\\

   \textbf{R$^2$}                    & 0.53738        &0.56883          & 0.55443          & 0.50716       &0.50702    & 0.50700 \\%Narrow: 0.53738 / 0.50269
   \textbf{Within R$^2$}              & 0.11473        & 0.13506          & 0.13132         & 0.13750       & 0.13725    & 0.13723\\%Narrow: 0.11473/ 0.14480
   \textbf{Fixed effects}            & \multicolumn{6}{c}{Bidder}  \\                         \bottomrule
\multicolumn{7}{l}{{\footnotesize Significance codes: ***: 0.001, **: 0.01, *: 0.05. Standard errors (clustered by bidder)}}\\
\multicolumn{7}{l}{{\footnotesize in parentheses. Bandwidths ISO-NE: $\pm 0.1$ (extra narrow), $\pm 0.2$ (polynomial), }}\\
\multicolumn{7}{l}{{\footnotesize $\pm0.5$ (medium). Bandwidths NYISO: $\pm 1.5$~\$/MWh (extra narrow), $\pm3$~\$/MWh}}\\
\multicolumn{7}{l}{{\footnotesize  (polynomial), $\pm 4.5$~\$/MWh (medium).}}\\\\
  \end{tabular}
\caption{Additional market-level regressions. The sign and magnitude of the treatment coefficient are moderately robust in ISO-NE, but not in NYISO. In both markets, adding polynomial coefficients weakens the explanatory power of the model.}
\label{tab:sensitivity}
\end{table}

\clearpage
\bibliographystyle{elsarticle-num} 
\bibliography{main}

\end{document}